\documentclass[a4paper,letters,fleqn,usenatbib]{mnras}

\usepackage{newtxtext,newtxmath}

\usepackage[T1]{fontenc}
\usepackage{ae,aecompl}

\usepackage{graphicx}	
\usepackage{amsmath}	
\usepackage{amssymb}	

\usepackage{rotating}
\usepackage{tikz}

\newcommand{\gaia}{\textit{Gaia}}
\newcommand{\gdrtwo}{\textit{Gaia}~DR2}

\title[The Galactic warp revealed by \emph{Gaia} DR2 kinematics]{The Galactic warp revealed by \emph{Gaia} DR2 kinematics}

\author[E. Poggio et al.]{
E.~Poggio,$^{1,2}$\thanks{E-mail: eloisa.poggio@inaf.it}
R.~Drimmel,$^{2}$
M.~G.~Lattanzi,$^{2}$
R.~L.~Smart,$^{2}$
A.~Spagna,$^{2}$ 
\newauthor
R.~Andrae,$^{3}$
C.~A.~L.~Bailer-Jones,$^{3}$
M.~Fouesneau,$^{3}$ 
T.~Antoja,$^{4}$
C.~Babusiaux,$^{5,7}$
\newauthor
D.~W.~Evans,$^{6}$
F.~Figueras,$^{4}$
D.~Katz,$^{7}$
C.~Reylé,$^{8}$
A.~C.~Robin,$^{8}$
\newauthor
M.~Romero-G\'{o}mez,$^{4}$
and G.~M.~Seabroke$^{9}$
\\
$^{1}$Universit\`a di Torino, Dipartimento di Fisica, via P. Giuria 1, I-10125, Torino, Italy\\
$^{2}$Osservatorio Astrofisico di Torino, Istituto Nazionale di Astrofisica (INAF), I-10025 Pino Torinese, Italy\\
$^{3}$Max Planck Institute for Astronomy, K\"onigstuhl 17, D-69117 Heidelberg, Germany\\
$^{4}$Institut de Ci\`{e}ncies del Cosmos, Universitat  de  Barcelona  (IEEC-UB), Mart\'{i} i Franqu\`{e}s  1, E-08028 Barcelona, Spain\\
$^{5}$Univ. Grenoble Alpes, CNRS, IPAG, 38000 Grenoble, France\\
$^{6}$Institute of Astronomy, University of Cambridge, Madingley Road, Cambridge CB3 0HA, UK\\
$^{7}$GEPI, Observatoire de Paris, Universit\'{e} PSL, CNRS, 5 Place Jules Janssen, 92190 Meudon, France\\
$^{8}$Institut UTINAM, CNRS UMR6213, Univ. Bourgogne Franche-Comt\'e, OSU THETA Franche-Comt\'e-Bourgogne, \\
Observatoire de Besan\c con, BP 1615, 25010 Besan\c con Cedex, France.\\
$^{9}$Mullard Space Science Laboratory, University College London, Holmbury St Mary, Dorking, Surrey RH5 6NT, United Kingdom\\
}

\date{Accepted XXX. Received YYY; in original form ZZZ}

\pubyear{2015}

\begin{document}
\label{firstpage}
\pagerange{\pageref{firstpage}--\pageref{lastpage}}
\maketitle

\begin{abstract}

Using \gdrtwo\, astrometry, we map the kinematic signature of the Galactic stellar warp out to a distance of 7 kpc from the Sun. Combining \gdrtwo\, and 2MASS photometry, we identify, via a probabilistic approach, $599\,494$ upper main sequence stars and $12\,616\,068$ giants without the need for individual extinction estimates. The spatial distribution of the upper main sequence stars clearly shows segments of the nearest spiral arms. The large-scale kinematics of both the upper main sequence and giant populations 
show a clear signature of the warp of the Milky Way, apparent as a gradient of 5-6 km/s in the vertical velocities from 8 to 14 kpc in Galactic radius. The presence of the signal in both samples, which have different typical ages, suggests that the warp is a gravitationally induced phenomenon.
%
%
\end{abstract}



\section{ Introduction }

The disc of our Galaxy was first seen to be warped in the radio observations of neutral hydrogen more than 60 years ago \citep{Kerr:1957}. Later observations 
\citep[][and others]{Freudenreich:1994,Drimmel:2001,LopezCorredoira:2002B,Robin:2008,Reyle:2009,Amores:2017} also showed that the stellar disc is flat out to roughly the Solar Circle, then bends up upwards in the north and downwards in the south, with the Sun close to the line of nodes. 
%
Theoretical models for the warping of stellar discs include interactions with satellites \citep{Kim:2014}, intergalactic magnetic fields \citep{Battaner:1990}, accretion of intergalactic matter \citep{Kahn:1959,LopezCorredoira:2002A}, and a mis-aligned dark halo \citep{Debattista:1999, Sparke:1988}, amongst others. However, to date only the shape of the Galactic warp has been roughly constrained, leading to a lack of consensus for its causal mechanism due to the lack of kinematic information perpendicular to the galactic disc. In particular, a consistent kinematic signature in old and young stars would exclude non-gravitational mechanisms (see Section \ref{Concl}).
In the pre-\gaia\ era, kinematics studies suggested a signature inconsistent with a long-lived warp \cite{Smart:1998,Drimmel:2000,LopezCorredoira:2014}$\,$, while the kinematics of stars near the Sun seemed to be consistent with the presence of a warp (\citealt{Dehnen:1998b}, though see \citealt{Seabroke:2007}). With the first \gaia\ data release, \cite{Schoenrich:2018} detected the warp kinematic signature using the TGAS catalogue, while \cite{WarpGaiaDR1:2017} found no evidence of the warp signal in the kinematics of OB stars.

With \gaia's most recent second data release, \cite{MWDR2:2018} (hereafter MWDR2) showed a kinematic signature on large scales consistent with a warp with a sample of red giants \citep[in agreement with LAMOST radial velocities,][]{Liu:2017}, while their young OB stellar sample seemed to give divergent results. In this contribution we expand on the work of MWDR2, with larger and fainter samples of the old (red giants) and young (upper main sequence stars) selected from \gdrtwo, using 2MASS \citep[2-Micron All Sky Survey,][]{2MASS:2006} photometry (Section \ref{SecData}). We compare the kinematic maps of these two samples (Section \ref{Kin}) and discuss the obtained results (Section \ref{Concl}).



\section{Data selection } \label{SecData}

\begin{figure}
\resizebox{\hsize}{!}{\includegraphics{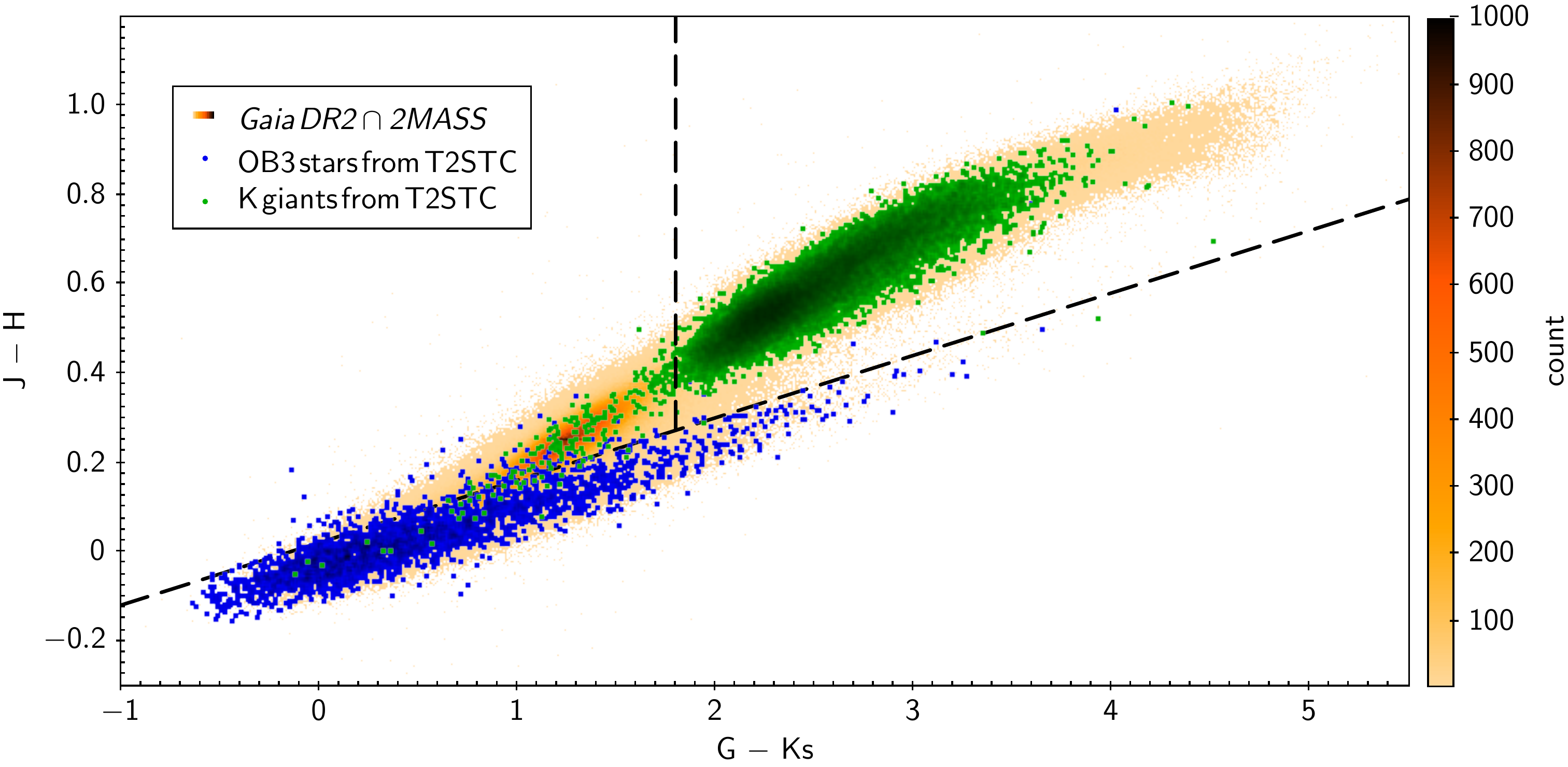}}
\caption{ Colour-colour plot showing the 2MASS-\gaia \, preliminary selection. Candidate upper main sequence (UMS) are taken as stars lying below the diagonal dashed line, while candidate giants are those lying in the top right area of the plots. A similar plot (here not shown) was constructed with $(J-K_s)$ on the vertical axis.
The yellow-orange density plots a sample of \gdrtwo\, stars with $G<12$, while the blue and green points show the colours of stars in the Tycho-2 Spectral Type Catalogue (T2STC) that are classified as either OB stars or K giants (luminosity class I and II). }
\label{colcolsel}
\end{figure}

To select upper main sequence (UMS) and giants in the Galactic plane ($|b|<20 \deg$) without the need for individual reddening estimates, we use 2MASS photometry for \gdrtwo \, sources using the cross match table provided by the \gaia \, Archive ({\footnotesize \texttt{https://archives.esac.esa.int/gaia}}), and restricting ourselves to 2MASS sources with uncertainties $\sigma_{J,H,K_s} < 0.05$ mag and a photometric quality flag of ``AAA". Finally, as a practical matter, we select stars with $G < 15.5$ mag, as very few fainter stars have 2MASS photometry. 

\begin{figure*}
\resizebox{\hsize}{!}{\includegraphics{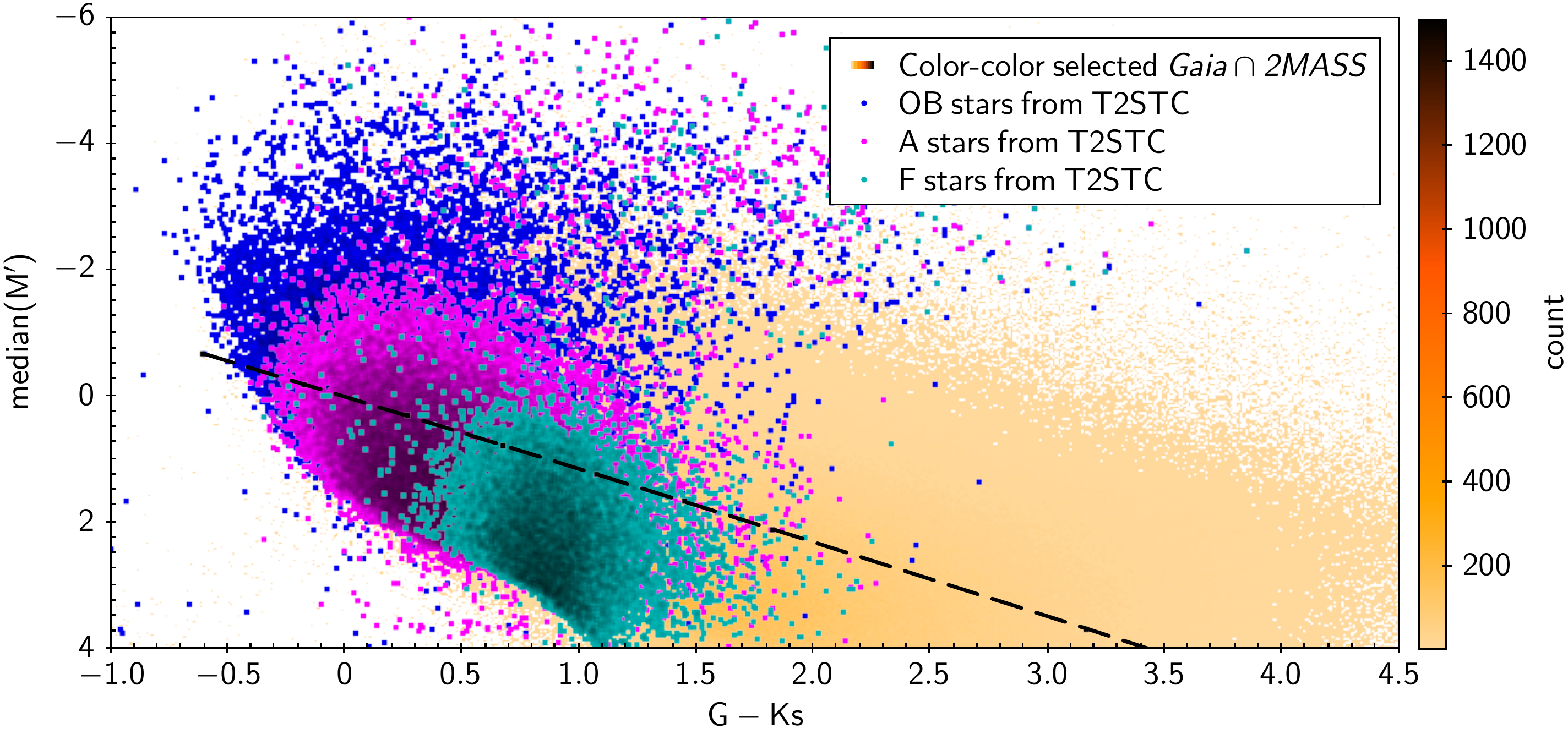} \quad \includegraphics{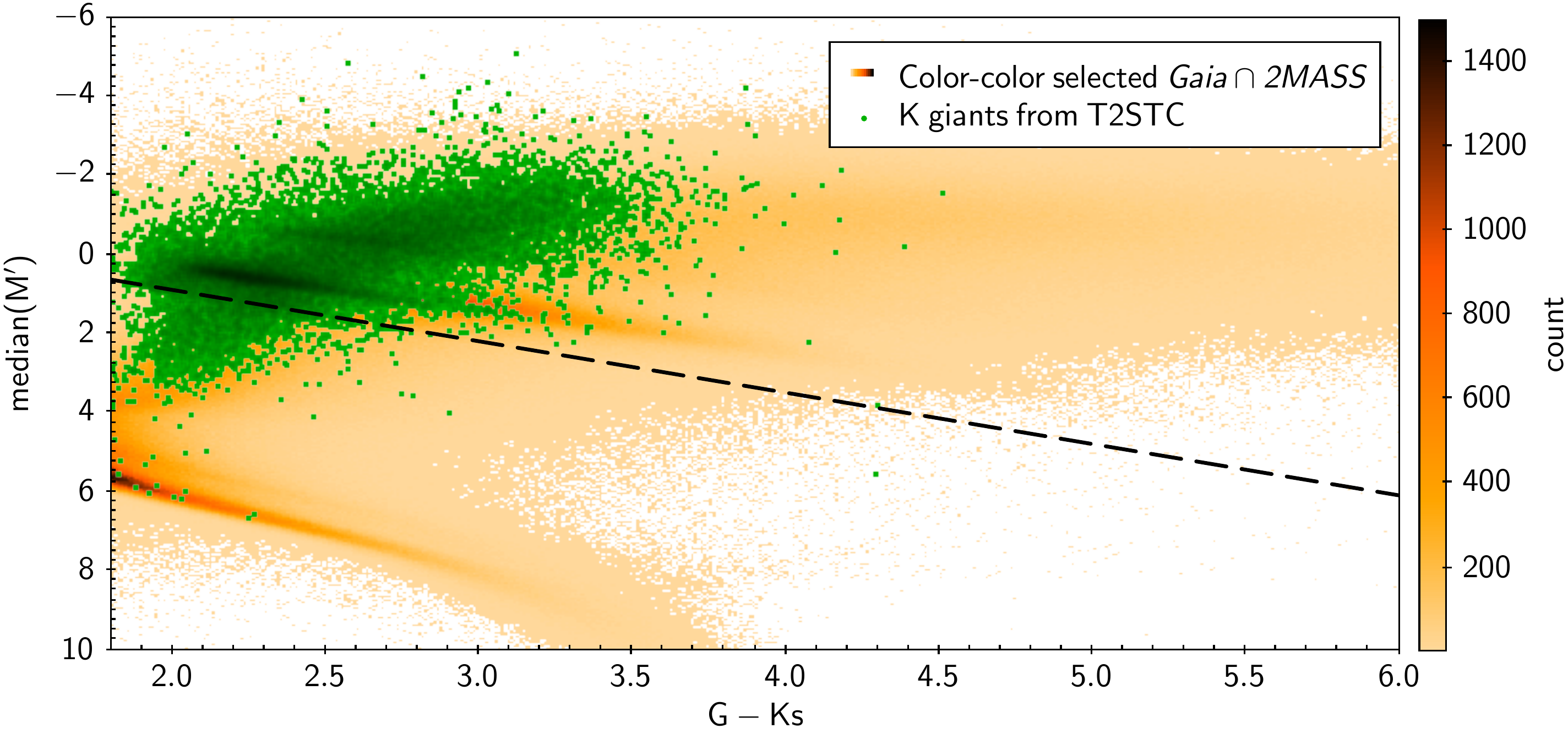}}
\caption{ The parallax criterium for the UMS (left panel) and for the giants (right panel). On the y-axis, the median of the probability density function of $M^\prime$. The dashed line shows the adopted tolerance limit (see text), selecting those stars that are above the dashed lines. Orange density area as in the previous plot, other coloured points are for those stars in the Tycho-2 Spectral Type Catalogue (T2STC), colour coded as per the key in the figures. }
\label{maglim}
\end{figure*}


{\bf  Upper Main Sequence stars.} A preliminary selection is made based only on measured 2MASS/\gaia \, colours. As shown in Figure \ref{colcolsel}, known OB stars from the Tycho-2 Spectral Type Catalogue \citep[hereafter T2STC,][]{T2STC:2003} lie along a sequence that is a consequence of interstellar reddening, which is clearly separated from the redder turn-off stars, giants and lower main sequence stars. Based on this, candidates UMS stars are selected from the Gaia DR2$\cap$2MASS catalogue satisfying both $(J-H) < 0.14 \, (G-K_s) + 0.02$ and $(J-K) < 0.23 \, (G-K_s)$. 

A second step of the selection procedure uses \gaia\, astrometry \citep{Lindegren:2018}, choosing those stars whose parallax $\varpi$, parallax uncertainty $\sigma_{\varpi}$ and apparent $G$ magnitude is likely to be consistent with being a UMS star. 
To this end, we calculated the probability density function (pdf) of the heliocentric distance $r$ for the given coordinates $(l,b)$ via Bayes' theorem, $P(r \, | l, b, \varpi, \sigma_{\varpi}) \propto P(\varpi|r,\sigma_{\varpi}) \, P(r | l, b) $, assuming a gaussian likelihood $P(\varpi|r,\sigma_{\varpi})$ and constructing the prior according to \cite{Astr:2016} (their Equation 7, i.e. the Milky Way prior) 
\setlength{\abovedisplayskip}{3pt}
\setlength{\belowdisplayskip}{3pt}
\begin {equation}
\label{prior_plx}
P(r|l,b) \propto r^2 \, \rho(l,b,r) \, S(l,b,r)  \quad .
\end{equation}
We adopt a simple density model for the Galactic disc $\rho(l,b,r)$, consisting of an exponential disc in Galactocentric radius R and vertical height $z$, with a radial scale length $L_{R} = 2.6$ kpc \citep{BlandGerhard:2016} and vertical scale height $h_z = 150$ pc \citep[larger than the known scale height for OB stars,][]{WarpGaiaDR1:2017}.
We assume for the Sun $R_{\odot}$=8.34 kpc \citep{Reid:2014} and $z_{\odot}=25$ pc \citep{BlandGerhard:2016}. The term $S(l,b,r)$ takes into account the fall-off of the number of observable objects with $r$ due to the survey selection function, neglect of which can cause severe biases in the obtained distance estimates \citep{Schoenrich:2017}. We estimated the term $S(l,b,r)$ according to \cite{Astr:2016}, and modelled the variation of Gaia DR2$\cap$2MASS completeness as a function of apparent magnitude G according to Drimmel et al. (in prep.), including the previously mentioned cut at $G=15.5$. The adopted luminosity function in the G band is calculated through the PARSEC isochrones \citep[web interface {\footnotesize \texttt{http://stev.oapd.inaf.it/cmd}},][]{Bressan:2012, Chen:2014,Chen:2015, Tang:2014}, after taking into account the colour-colour cuts applied in the preliminary selection. The luminosity function was obtained assuming a star formation rate constant with time, the canonical two-part power law IMF corrected for unresolved binaries \citep[][]{Kroupa:2001, Kroupa:2002}, and solar metallicity. 
The impact of various assumptions incorporated in our prior is discussed in the following section. 

For each star, we derived from $P(r \, | l, b, \varpi, \sigma_{\varpi})$ a pdf of the quantity $M^\prime \equiv M_G + A_G = \ G - 5 \log{r_{pc}} + 5$, which is the absolute magnitude $M_G$ plus the extinction in the $G$ band of the source. The Jacobian of the transformation $d r /d M^\prime$ can be written when the $G$ magnitude is fixed, obtaining $P( M^\prime | l, b, \varpi, \sigma_{\varpi}, G)$. After numerically imposing the normalization condition $\int^{+\infty}_{-\infty} P(M^\prime | l, b, \varpi, \sigma_{\varpi}, G) d M^\prime = 1$, we calculate the probability of the star being brighter than the limit $M^\prime_{lim}$, which is the faintest extincted magnitude that we are willing to tolerate for an UMS star candidate with an observed $(G-K_s)$ colour. The tolerance limit $M^\prime_{lim}$ was arbitrarily chosen as the absolute magnitude of a fictitious B3-like star having $\log( \text{age/yr})=6$ and $\log (T_{eff}) = 4.27$. For such a star, the PARSEC isochrones provide us with an absolute magnitude of $(M_G)_{lim}=-0.7$ and $(G-K_s)=-0.6$ in the case of no extinction. The PARSEC isochrones give the corresponding values of $M^\prime_{lim}$ and $(G-K_s)$ when extinction is present (Figure \ref{maglim}, left plot). 
Hence we calculate the probability of the star being an UMS star - i.e. brighter than the tolerance limit - by performing the following integral 
\setlength{\abovedisplayskip}{3pt}
\setlength{\belowdisplayskip}{3pt}
\begin {equation}
\label{prob_OB}
p( UMS \, |\, l,  b, \varpi, \sigma_{\varpi}, G) \! = \!  \int_{-\infty}^{M^\prime_{lim}} P( M^\prime | l, b, \varpi, \sigma_{\varpi}, G) \, d M^\prime ,
\end{equation}
which is by definition between 0 and 1. The stars for which $p ( UMS \, | \, l,  b, \varpi, \sigma_{\varpi}, G) > 0.5$ are selected, giving us 599\,494 UMS stars. 


{\bf Giant stars.} In a similar fashion as the colour-colour selection of the UMS stars, we perform a preliminary selection based on photometry, this time selecting the stars with $(J-H) > 0.14\,(G-Ks) + 0.02$ and $(J-Ks) > 0.23\,(G-Ks)$ (see Figure \ref{colcolsel}), with an additional $(G-Ks) > 1.8$ cut to remove the objects too blue to be considered giant candidates. We adopt the same probabilistic approach used for the UMS stars, but assuming a spatial density scale height of $h_z=$300 pc \citep{BlandGerhard:2016}. We calculate for each source the probability of being a giant star, with the tolerance limit set as equal to $M^\prime_{lim}=1.3 \, (G-Ks) - 1.7$. Such a limit removes sub-giants and dwarfs, and also accounts for interstellar reddening (see Figure \ref{maglim}, right plot). This selection gives us 12\,616\,068 giants. 


To test the composition of the selected samples, we crossmatched our samples with the T2STC. For the UMS sample, we obtained 24\,422 objects, of which approximately $55\%$ are OB stars, $40\%$ are A stars and $5\%$ are F stars, according to the T2STC spectral classifications. For the giant sample, we found 33\,842 stars with complete spectral classification from T2STC, of which $88\%$ are giants ($69\%$ K giants and $19\%$ G giants) and $12\%$ are main sequence stars (mostly of spectral class K or G, while A or F stars are less than 1\%).

\vspace*{-\baselineskip}

\section{Density and kinematic maps }\label{Kin}

\begin{figure*}
\centering
\includegraphics[clip=true, trim = 10mm 60mm 10mm 60mm, width=0.41  \hsize]{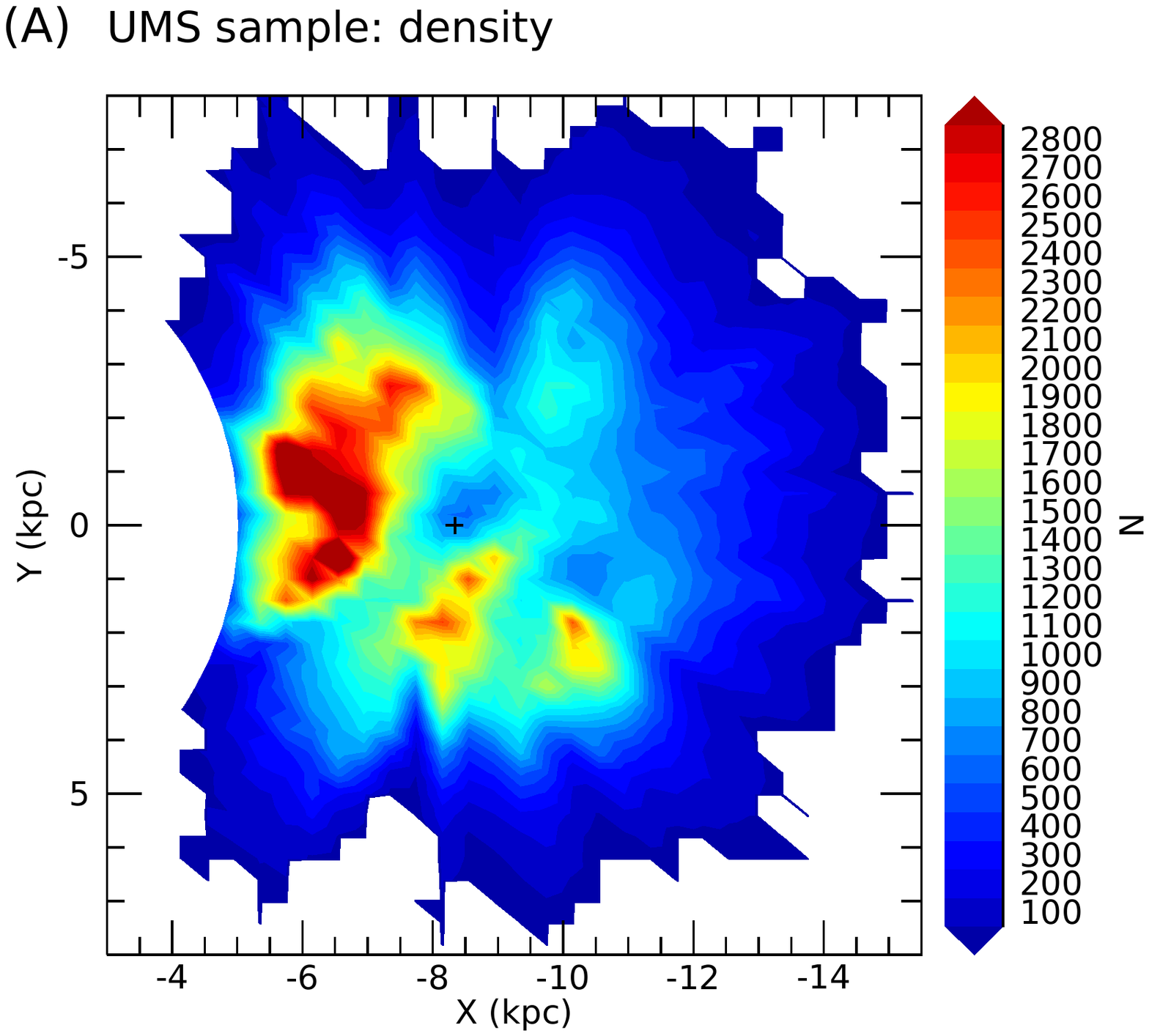}
\includegraphics[clip=true, trim = 10mm 60mm 10mm 60mm, width=0.41 \hsize]{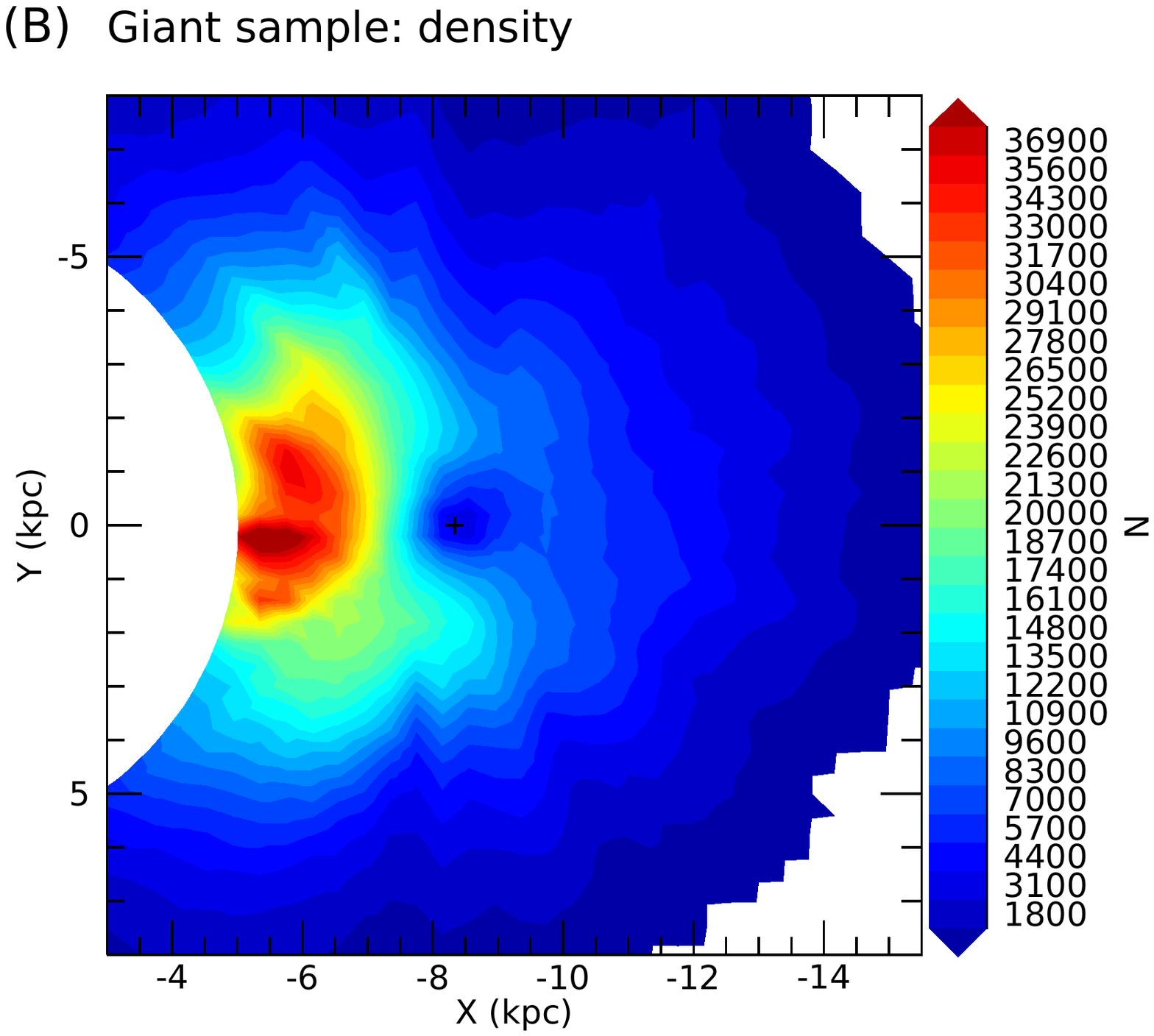}
\includegraphics[clip=true, trim = 10mm 60mm 10mm 60mm, width=0.41  \hsize]{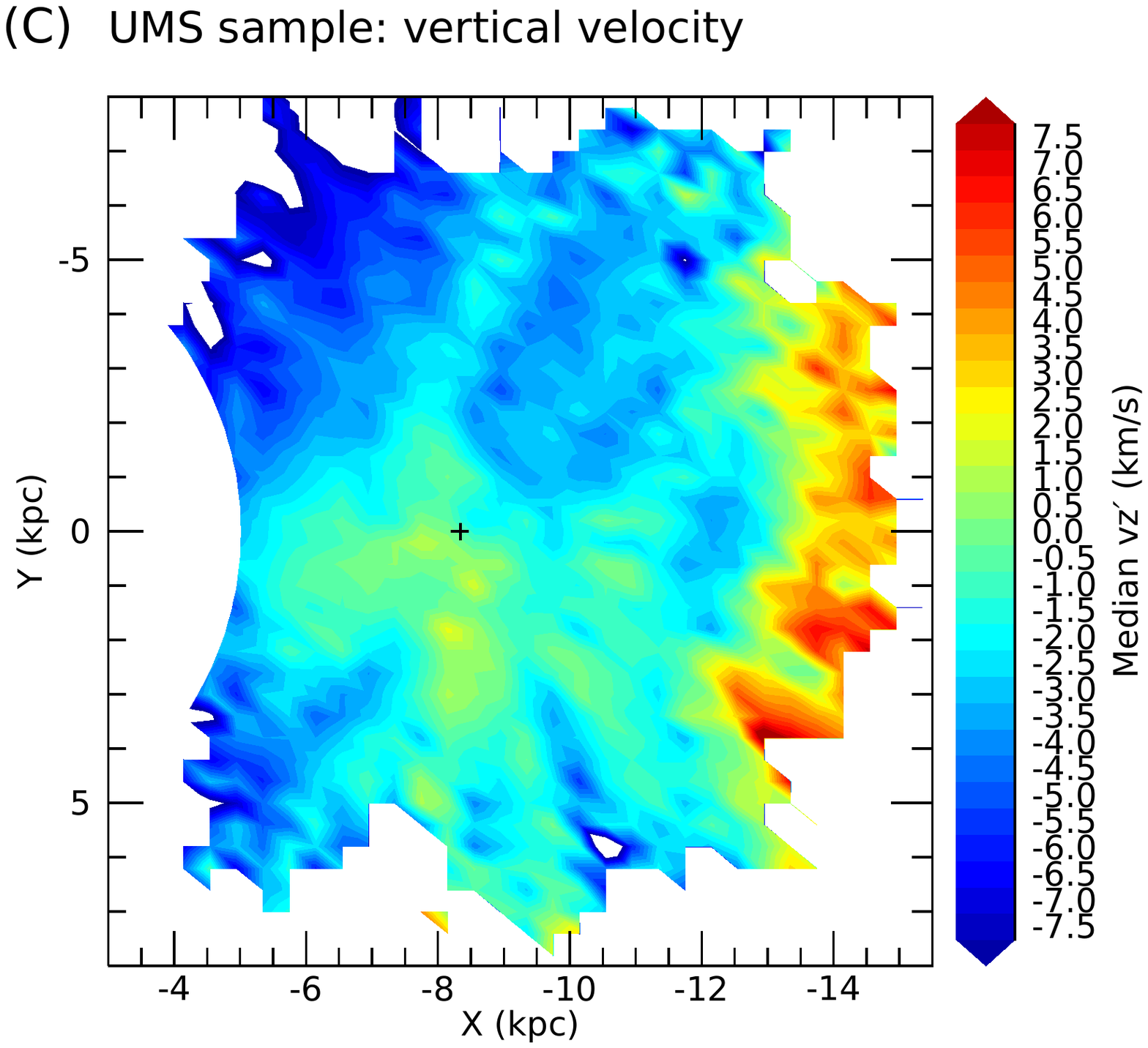}
\includegraphics[clip=true, trim = 10mm 60mm 10mm 60mm, width=0.41  \hsize]{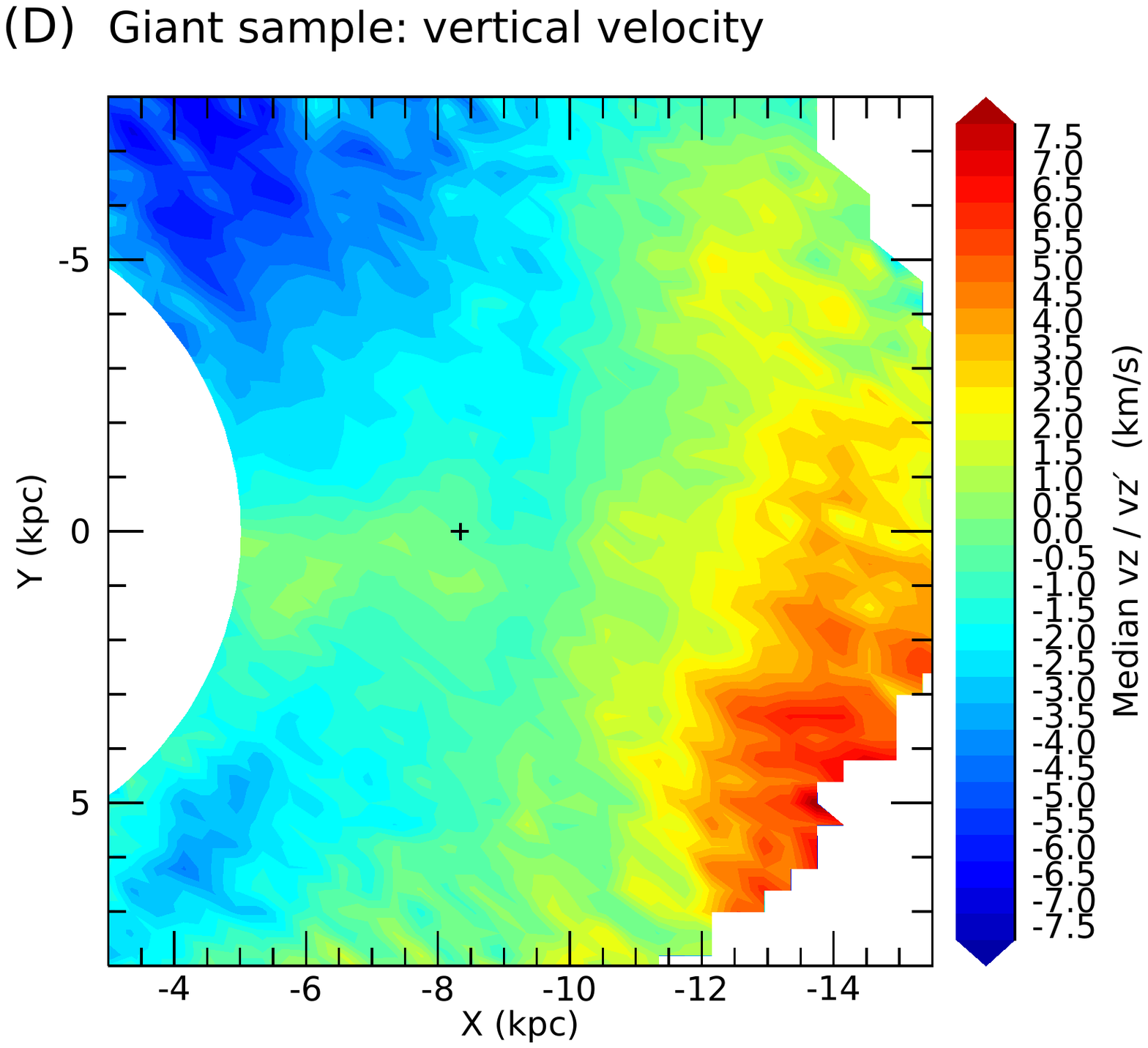}
\includegraphics[clip=true, trim = 10mm 60mm 10mm 60mm, width=0.41  \hsize]{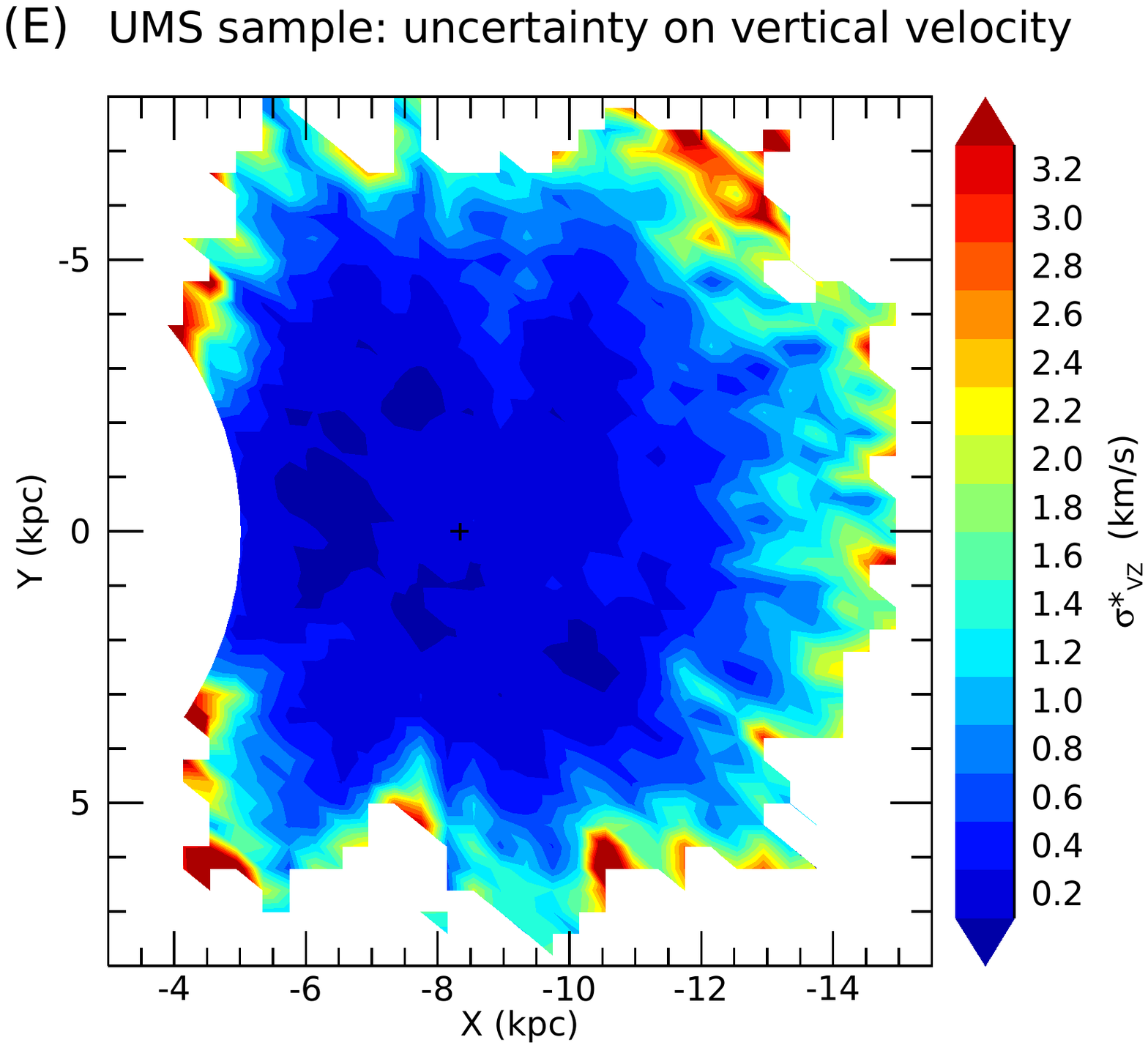}
\includegraphics[clip=true, trim = 10mm 60mm 10mm 60mm, width=0.41  \hsize]{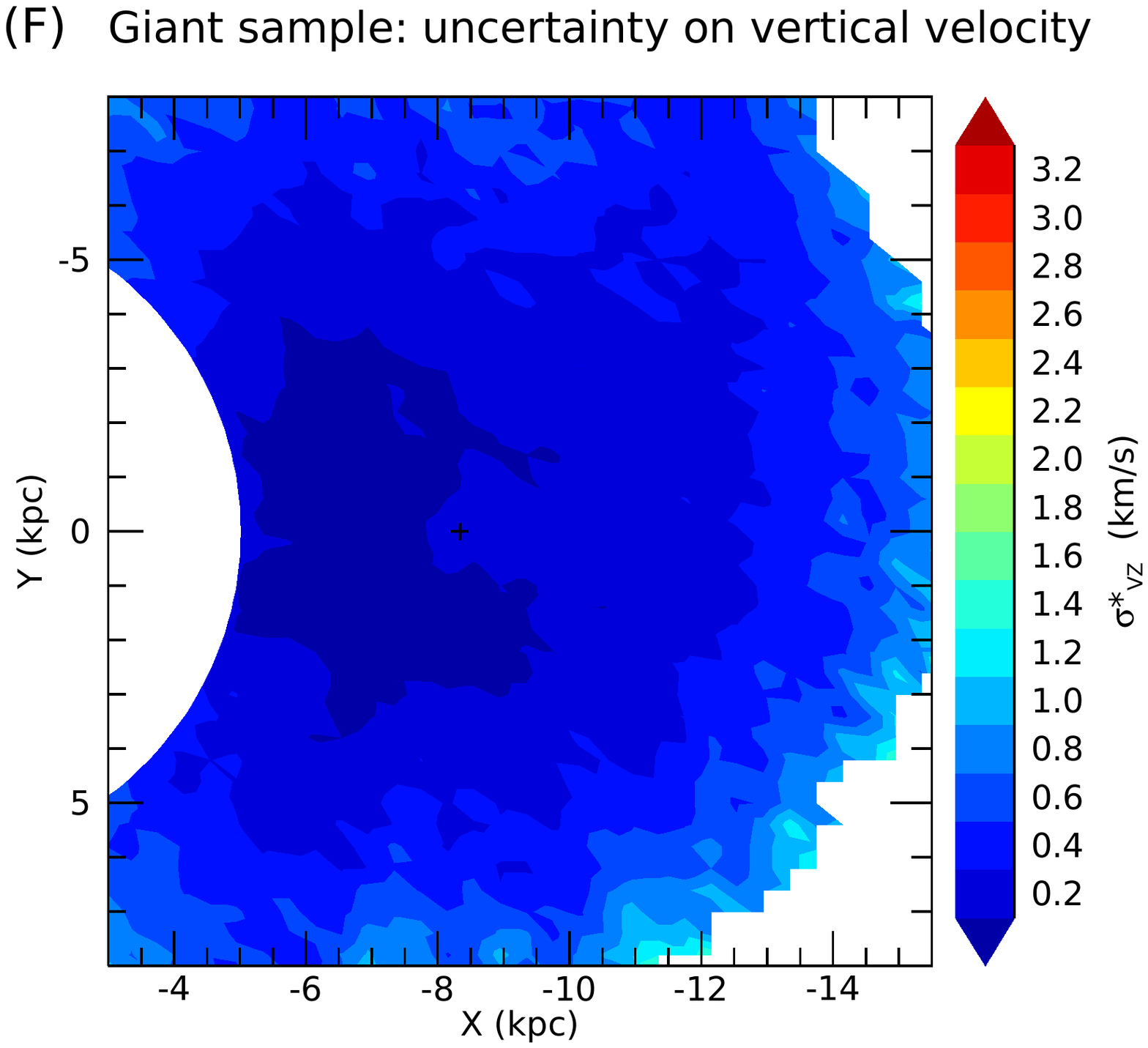}
 \caption{Maps for the UMS (left plots) and giant (right plots) samples. The Sun is represented by a black cross at X = -8.35 kpc and Y=0 kpc. The Galactic center is located at X=0 and Y=0, and the Galaxy is rotating clockwise. The XY plane was divided into cells of 400 pc width, only showing the ones containing more than 50 /500 stars for the UMS/giant sample. From top to bottom: maps of the density (N is the number of sources per cell), median vertical velocity $V_Z$ or $V^{\prime}_Z$ (see text) and bootstrap uncertainty on the median vertical velocity $\sigma^*_{V_Z}$. 
 }\label{kinmaps}
\end{figure*}

In this Section, we present and compare the maps obtained with the UMS and giant samples, shown in Figure \ref{kinmaps}. For both samples 
we use as our distance estimator for each star the mean \citep[see for example][]{MacKay:2003,Gelman:1995aa} of the posterior distribution $P(r \, | l, b, \varpi, \sigma_{\varpi})$ (see previous section). 
The UMS stars have mean distances of approximately 3 kpc, and mean heights with respect to the Galactic plane of about 100 pc, in contrast to the giant sample, which presents, respectively, 4.5 kpc and 480 pc.
The giant sample exhibits a smooth density distribution (Figure \ref{kinmaps}B), decreasing for large heliocentric distance, as expected for a magnitude limited sample, and for larger Galactocentric radii, as expected from an exponential disc. In contrast the UMS sample (Figure \ref{kinmaps}A) shows three observed overdensities that correspond to sections of the nearby spiral arms (from left to right: Sagittarius-Carina arm, local arm and Perseus arm). The evident spiral structure confirms that our UMS sample is young with respect to the smooth distribution shown by the older and dynamically relaxed giant population. 


Figure \ref{kinmaps}C and \ref{kinmaps}D show a face-on view of the vertical motions in the Galactic plane of the two samples, calculated deriving the proper motions in galactic latitude $\mu_b$ from the \gdrtwo\, astrometry and correcting for the solar motion $(V_{X\odot},V_{Y\odot},V_{Z\odot})=(11.1, 12.24, 7.25)$ km s$^{-1}$ \citep{Schoenrich:2010}. The large majority of stars in our UMS sample 
lack line-of-sight velocities, so that it is not possible to calculate directly the vertical velocity. We therefore estimate the mean vertical velocity $V^\prime_Z$ from the available astrometry, correcting for solar motion and differential Galactic rotation, assuming a flat rotation curve \citep[$V_c = 240$ km/s, ][]{Reid:2014}, as done in MWDR2 \citep[see Equation 8 of][]{Drimmel:2000}.
We find that $3\,042\,265$ of our giants have line-of-sight velocities provided in \gdrtwo, for which we calculate directly the vertical velocity, while for the remaining we estimate the vertical velocities as done for the UMS sample. (For the subsample of star having line-of-sight velocities, we have verified that our approximation of using $V^\prime_Z$ instead of $V_Z$ produces consistent results.)

A gradient in the median vertical velocities is apparent in Figure \ref{kinmaps}C and \ref{kinmaps}D, as expected from a warp signature \citep{Abedi:2014,WarpGaiaDR1:2017}. Also worthy to note is that the peak velocities in both
samples is not exactly toward the anti-center, which is probably due to the Sun not being on the line-of-nodes. Radial features in this plot are due to uneven sampling above/below the Galactic plane due to foreground extinction (see Section 8.4.2 in the Gaia DR2 online documentation). The bootstrap uncertainties on the median velocities $\sigma^*_{V_Z}$ are shown in Figure \ref{kinmaps}E and \ref{kinmaps}F. The systematic increase of the median vertical velocity is of about 5-6 km/s from $R \sim 8$ kpc to $14$ kpc, with a signal-to-noise greater than 10. The subsets of stars having $\varpi/\sigma_{\varpi}>5$ ($478\,258$ UMS stars and $6\,373\,188$ giants) present a signal consistent with the whole sample. In order to test the robustness of the signal, we also re-calculated distances with the iterative approach of \cite{Schoenrich:2017} for $20^o<l<340^o$, finding a consistent gradient. We also slightly modified the prior (e.g. assuming $L_R=4$ kpc for the UMS sample or including a thick disc for the giant sample), always confirming the presence of the signal. Moreover, we verified that adopting as distance estimator the mode \citep[following][]{BailerJones:2018,BailerJones:2017,BailerJones:2015} or the median of the pdf produces consistent results. Finally, we explored the impact of a systematic zero-point error (exploring the range $\pm 0.080$ mas) of \gdrtwo\ parallaxes \citep{Lindegren:2018}, which only results in a contraction/expansion of the maps, but still preserves the presence of the warp signature.


\section{Discussion and conclusions}  \label{Concl}

The kinematic signature of the Galactic warp is expected to manifest itself toward the Galactic anticenter as large-scale systematic velocities perpendicular to the Galactic plane. Thanks to the large sample of stars in \gdrtwo\ with exquisite astrometric precision, we are able to map the vertical motions over a larger extent of the Milky Way's disc than previously possible, for both an intrinsically young and old population. That our UMS sample clearly shows the spiral arms, in contrast with the giant population, confirms that it is a dynamically young population.

The observed gradient in the giants appears to be in agreement with the overall increase in vertical velocity shown by \gdrtwo\ data in \cite{Kawata:2018} and the giant sample in MWDR2 for the range in Galactocentric radius in which our studies overlap. Meanwhile, our UMS sample exhibit a more perturbed pattern than the giants at $R<12$ kpc, in agreement with the OB sample in MWDR2, showing the warp signature at larger Galactocentric radii.

The presence of the warp signature in our two samples suggests that the warp is principally a gravitational phenomenon; indeed, warp generation models exclusively based on non-gravitational mechanisms (such as magnetic fields or hydrodynamical pressure from infalling gas) would act on the gas and affect the young stars only \citep[see also the discussion in][]{Guijarro:2010,Sellwood:2013}; recently-born stars would inherit the kinematics of the gas and trace the warp-induced kinematics until phase mixing smeared out evidence of their initial conditions. The detection of a similar warp kinematic signal in both young and old stellar populations thus suggests that gravity is the principle mechanism causing the warp. However, the two samples do present some differences on smaller scales, possibly indicating that additional perturbations or forces are acting on the gaseous component of the disc.


We have here only evidenced the kinematic signature of the warp in \gdrtwo.  Our findings bear further witness to the great potential of this data set. Future work confronting this signature with more quantitative models will certainly reveal further details of the dynamical nature of the Galactic warp.


\section*{Acknowledgements}

This work makes use of data products from: the ESA \gaia\
mission (\url{gea.esac.esa.int/archive/}), funded by national institutions
participating in the \gaia~Multilateral Agreement; and the Two Micron All Sky Survey (2MASS,
\url{www.ipac.caltech.edu/2mass}). This work was supported by ASI (Italian Space Agency) under contract 2014-025-R.1.2015 and the MINECO (Spanish Ministry of Economy) through grant ESP2016-80079-C2-1-R (MINECO/FEDER, UE) and MDM-2014-0369 of ICCUB (Unidad de Excelencia 'María de Maeztu'). EP acknowledges the financial support of the 2014 PhD fellowship programme of the INAF (Istituto Nazionale di Astrofisica). 
This project was developed in part at the 2016 NYC Gaia Sprint, hosted by the Center for Computational Astrophysics at the Simons Foundation in New York City.  




\vspace*{-\baselineskip}

\bibliographystyle{mnras}
\bibliography{mybib2} 

\begin{thebibliography}{}
\makeatletter
\relax
\def\mn@urlcharsother{\let\do\@makeother \do\$\do\&\do\#\do\^\do\_\do\%\do\~}
\def\mn@doi{\begingroup\mn@urlcharsother \@ifnextchar [ {\mn@doi@}
  {\mn@doi@[]}}
\def\mn@doi@[#1]#2{\def\@tempa{#1}\ifx\@tempa\@empty \href
  {http://dx.doi.org/#2} {doi:#2}\else \href {http://dx.doi.org/#2} {#1}\fi
  \endgroup}
\def\mn@eprint#1#2{\mn@eprint@#1:#2::\@nil}
\def\mn@eprint@arXiv#1{\href {http://arxiv.org/abs/#1} {{\tt arXiv:#1}}}
\def\mn@eprint@dblp#1{\href {http://dblp.uni-trier.de/rec/bibtex/#1.xml}
  {dblp:#1}}
\def\mn@eprint@#1:#2:#3:#4\@nil{\def\@tempa {#1}\def\@tempb {#2}\def\@tempc
  {#3}\ifx \@tempc \@empty \let \@tempc \@tempb \let \@tempb \@tempa \fi \ifx
  \@tempb \@empty \def\@tempb {arXiv}\fi \@ifundefined
  {mn@eprint@\@tempb}{\@tempb:\@tempc}{\expandafter \expandafter \csname
  mn@eprint@\@tempb\endcsname \expandafter{\@tempc}}}

\bibitem[\protect\citeauthoryear{{Abedi}, {Mateu}, {Aguilar}, {Figueras}  \&
  {Romero-G{\'o}mez}}{{Abedi} et~al.}{2014}]{Abedi:2014}
{Abedi} H.,  {Mateu} C.,  {Aguilar} L.~A.,  {Figueras} F.,   {Romero-G{\'o}mez}
  M.,  2014, \mn@doi [\mnras] {10.1093/mnras/stu1035}, \href
  {http://adsabs.harvard.edu/abs/2014MNRAS.442.3627A} {442, 3627}

\bibitem[\protect\citeauthoryear{{Am{\^o}res}, {Robin}  \&
  {Reyl{\'e}}}{{Am{\^o}res} et~al.}{2017}]{Amores:2017}
{Am{\^o}res} E.~B.,  {Robin} A.~C.,   {Reyl{\'e}} C.,  2017, \mn@doi [\aap]
  {10.1051/0004-6361/201628461}, \href
  {http://adsabs.harvard.edu/abs/2017A%26A...602A..67A} {602, A67}

\bibitem[\protect\citeauthoryear{{Astraatmadja} \&
  {Bailer-Jones}}{{Astraatmadja} \& {Bailer-Jones}}{2016}]{Astr:2016}
{Astraatmadja} T.~L.,  {Bailer-Jones} C.~A.~L.,  2016, \mn@doi [\apj]
  {10.3847/0004-637X/832/2/137}, \href
  {http://adsabs.harvard.edu/abs/2016ApJ...832..137A} {832, 137}

\bibitem[\protect\citeauthoryear{{Bailer-Jones}}{{Bailer-Jones}}{2015}]{BailerJones:2015}
{Bailer-Jones} C.~A.~L.,  2015, \mn@doi [\pasp] {10.1086/683116}, \href
  {http://adsabs.harvard.edu/abs/2015PASP..127..994B} {127, 994}

\bibitem[\protect\citeauthoryear{Bailer-Jones}{Bailer-Jones}{2017}]{BailerJones:2017}
Bailer-Jones C. A.~L.,  2017, Practical Bayesian Inference: A Primer for
  Physical Scientists.
Cambridge University Press, \mn@doi{10.1017/9781108123891}

\bibitem[\protect\citeauthoryear{{Bailer-Jones}, {Rybizki}, {Fouesneau},
  {Mantelet}  \& {Andrae}}{{Bailer-Jones} et~al.}{2018}]{BailerJones:2018}
{Bailer-Jones} C.~A.~L.,  {Rybizki} J.,  {Fouesneau} M.,  {Mantelet} G.,
  {Andrae} R.,  2018, preprint, \href
  {http://adsabs.harvard.edu/abs/2018arXiv180410121B} {} (\mn@eprint {AJ, in
  press, arXiv} {1804.10121})

\bibitem[\protect\citeauthoryear{{Battaner}}{{Battaner}}{1990}]{Battaner:1990}
{Battaner} E. e.~a.,  1990, \aap, \href
  {http://adsabs.harvard.edu/abs/1990A%26A...236....1B} {236, 1}

\bibitem[\protect\citeauthoryear{{Bland-Hawthorn} \&
  {Gerhard}}{{Bland-Hawthorn} \& {Gerhard}}{2016}]{BlandGerhard:2016}
{Bland-Hawthorn} J.,  {Gerhard} O.,  2016, \mn@doi [\araa]
  {10.1146/annurev-astro-081915-023441}, \href
  {http://adsabs.harvard.edu/abs/2016ARA%26A..54..529B} {54, 529}

\bibitem[\protect\citeauthoryear{{Bressan}, {Marigo}, {Girardi}, {Salasnich},
  {Dal Cero}, {Rubele}  \& {Nanni}}{{Bressan} et~al.}{2012}]{Bressan:2012}
{Bressan} A.,  {Marigo} P.,  {Girardi} L.,  {Salasnich} B.,  {Dal Cero} C.,
  {Rubele} S.,   {Nanni} A.,  2012, \mn@doi [\mnras]
  {10.1111/j.1365-2966.2012.21948.x}, \href
  {http://adsabs.harvard.edu/abs/2012MNRAS.427..127B} {427, 127}

\bibitem[\protect\citeauthoryear{{Chen}, {Girardi}, {Bressan}, {Marigo},
  {Barbieri}  \& {Kong}}{{Chen} et~al.}{2014}]{Chen:2014}
{Chen} Y.,  {Girardi} L.,  {Bressan} A.,  {Marigo} P.,  {Barbieri} M.,   {Kong}
  X.,  2014, \mn@doi [\mnras] {10.1093/mnras/stu1605}, \href
  {http://adsabs.harvard.edu/abs/2014MNRAS.444.2525C} {444, 2525}

\bibitem[\protect\citeauthoryear{{Chen}, {Bressan}, {Girardi}, {Marigo}, {Kong}
   \& {Lanza}}{{Chen} et~al.}{2015}]{Chen:2015}
{Chen} Y.,  {Bressan} A.,  {Girardi} L.,  {Marigo} P.,  {Kong} X.,   {Lanza}
  A.,  2015, \mn@doi [\mnras] {10.1093/mnras/stv1281}, \href
  {http://adsabs.harvard.edu/abs/2015MNRAS.452.1068C} {452, 1068}

\bibitem[\protect\citeauthoryear{{Debattista} \& {Sellwood}}{{Debattista} \&
  {Sellwood}}{1999}]{Debattista:1999}
{Debattista} V.~P.,  {Sellwood} J.~A.,  1999, \mn@doi [\apjl] {10.1086/311913},
  \href {http://adsabs.harvard.edu/abs/1999ApJ...513L.107D} {513, L107}

\bibitem[\protect\citeauthoryear{{Dehnen}}{{Dehnen}}{1998}]{Dehnen:1998b}
{Dehnen} W.,  1998, \mn@doi [\aj] {10.1086/300364}, \href
  {http://adsabs.harvard.edu/abs/1998AJ....115.2384D} {115, 2384}

\bibitem[\protect\citeauthoryear{{Drimmel} \& {Spergel}}{{Drimmel} \&
  {Spergel}}{2001}]{Drimmel:2001}
{Drimmel} R.,  {Spergel} D.~N.,  2001, \mn@doi [\apj] {10.1086/321556}, \href
  {http://adsabs.harvard.edu/abs/2001ApJ...556..181D} {556, 181}

\bibitem[\protect\citeauthoryear{{Drimmel}, {Smart}  \& {Lattanzi}}{{Drimmel}
  et~al.}{2000}]{Drimmel:2000}
{Drimmel} R.,  {Smart} R.~L.,   {Lattanzi} M.~G.,  2000, \aap, \href
  {http://adsabs.harvard.edu/abs/2000A%26A...354...67D} {354, 67}

\bibitem[\protect\citeauthoryear{{Freudenreich} et~al.,}{{Freudenreich}
  et~al.}{1994}]{Freudenreich:1994}
{Freudenreich} H.~T.,  et~al., 1994, \mn@doi [\apjl] {10.1086/187415}, \href
  {http://adsabs.harvard.edu/abs/1994ApJ...429L..69F} {429, L69}

\bibitem[\protect\citeauthoryear{{Gaia Collaboration} et~al.,}{{Gaia
  Collaboration} et~al.}{2018}]{MWDR2:2018}
{Gaia Collaboration} et~al., 2018, preprint, \href
  {http://adsabs.harvard.edu/abs/2018arXiv180409380G} {} (\mn@eprint {MWDR2, in
  press, arXiv} {1804.09380})

\bibitem[\protect\citeauthoryear{Gelman, Gelman, Robert, Chopin  \&
  Rousseau}{Gelman et~al.}{1995}]{Gelman:1995aa}
Gelman A.,  Gelman A.,  Robert C.,  Chopin N.,   Rousseau J.,  1995, ]
  {10.1.1.217.2021}, pp 1360--1383

\bibitem[\protect\citeauthoryear{{Guijarro}, {Peletier}, {Battaner},
  {Jim{\'e}nez-Vicente}, {de Grijs}  \& {Florido}}{{Guijarro}
  et~al.}{2010}]{Guijarro:2010}
{Guijarro} A.,  {Peletier} R.~F.,  {Battaner} E.,  {Jim{\'e}nez-Vicente} J.,
  {de Grijs} R.,   {Florido} E.,  2010, \mn@doi [\aap]
  {10.1051/0004-6361/201014506}, \href
  {http://adsabs.harvard.edu/abs/2010A%26A...519A..53G} {519, A53}

\bibitem[\protect\citeauthoryear{{Kahn} \& {Woltjer}}{{Kahn} \&
  {Woltjer}}{1959}]{Kahn:1959}
{Kahn} F.~D.,  {Woltjer} L.,  1959, \mn@doi [\apj] {10.1086/146762}, \href
  {http://adsabs.harvard.edu/abs/1959ApJ...130..705K} {130, 705}

\bibitem[\protect\citeauthoryear{{Kawata}, {Baba}, {Ciuc{\v a}}, {Cropper},
  {Grand}, {Hunt}  \& {Seabroke}}{{Kawata} et~al.}{2018}]{Kawata:2018}
{Kawata} D.,  {Baba} J.,  {Ciuc{\v a}} I.,  {Cropper} M.,  {Grand} R.~J.~J.,
  {Hunt} J.~A.~S.,   {Seabroke} G.,  2018, \mn@doi [\mnras]
  {10.1093/mnrasl/sly107}, \href
  {http://adsabs.harvard.edu/abs/2018MNRAS.479L.108K} {479, L108}

\bibitem[\protect\citeauthoryear{{Kerr}}{{Kerr}}{1957}]{Kerr:1957}
{Kerr} F.~J.,  1957, \mn@doi [\aj] {10.1086/107466}, \href
  {http://adsabs.harvard.edu/abs/1957AJ.....62...93K} {62, 93}

\bibitem[\protect\citeauthoryear{{Kim}, {Peirani}, {Kim}, {Ann}, {An}  \&
  {Yoon}}{{Kim} et~al.}{2014}]{Kim:2014}
{Kim} J.~H.,  {Peirani} S.,  {Kim} S.,  {Ann} H.~B.,  {An} S.-H.,   {Yoon}
  S.-J.,  2014, \mn@doi [\apj] {10.1088/0004-637X/789/1/90}, \href
  {http://adsabs.harvard.edu/abs/2014ApJ...789...90K} {789, 90}

\bibitem[\protect\citeauthoryear{{Kroupa}}{{Kroupa}}{2001}]{Kroupa:2001}
{Kroupa} P.,  2001, \mn@doi [\mnras] {10.1046/j.1365-8711.2001.04022.x}, \href
  {http://adsabs.harvard.edu/abs/2001MNRAS.322..231K} {322, 231}

\bibitem[\protect\citeauthoryear{{Kroupa}}{{Kroupa}}{2002}]{Kroupa:2002}
{Kroupa} P.,  2002, \mn@doi [Science] {10.1126/science.1067524}, \href
  {http://adsabs.harvard.edu/abs/2002Sci...295...82K} {295, 82}

\bibitem[\protect\citeauthoryear{{Lindegren} et~al.,}{{Lindegren}
  et~al.}{2018}]{Lindegren:2018}
{Lindegren} L.,  et~al., 2018, preprint, \href
  {http://adsabs.harvard.edu/abs/2018arXiv180409366L} {} (\mn@eprint {in press,
  arXiv} {1804.09366})

\bibitem[\protect\citeauthoryear{{Liu}, {Tian}  \& {Wan}}{{Liu}
  et~al.}{2017}]{Liu:2017}
{Liu} C.,  {Tian} H.-J.,   {Wan} J.-C.,  2017. pp~6--9 (\mn@eprint {arXiv}
  {1702.02233}), \mn@doi{10.1017/S1743921316011029}

\bibitem[\protect\citeauthoryear{{L{\'o}pez-Corredoira}, {Betancort-Rijo}  \&
  {Beckman}}{{L{\'o}pez-Corredoira} et~al.}{2002a}]{LopezCorredoira:2002A}
{L{\'o}pez-Corredoira} M.,  {Betancort-Rijo} J.,   {Beckman} J.~E.,  2002a,
  \mn@doi [\aap] {10.1051/0004-6361:20020229}, \href
  {http://adsabs.harvard.edu/abs/2002A%26A...386..169L} {386, 169}

\bibitem[\protect\citeauthoryear{{L{\'o}pez-Corredoira}, {Cabrera-Lavers},
  {Garz{\'o}n}  \& {Hammersley}}{{L{\'o}pez-Corredoira}
  et~al.}{2002b}]{LopezCorredoira:2002B}
{L{\'o}pez-Corredoira} M.,  {Cabrera-Lavers} A.,  {Garz{\'o}n} F.,
  {Hammersley} P.~L.,  2002b, \mn@doi [\aap] {10.1051/0004-6361:20021175},
  \href {http://adsabs.harvard.edu/abs/2002A%26A...394..883L} {394, 883}

\bibitem[\protect\citeauthoryear{{L{\'o}pez-Corredoira}, {Abedi}, {Garz{\'o}n}
  \& {Figueras}}{{L{\'o}pez-Corredoira} et~al.}{2014}]{LopezCorredoira:2014}
{L{\'o}pez-Corredoira} M.,  {Abedi} H.,  {Garz{\'o}n} F.,   {Figueras} F.,
  2014, \mn@doi [\aap] {10.1051/0004-6361/201424573}, \href
  {http://adsabs.harvard.edu/abs/2014A%26A...572A.101L} {572, A101}

\bibitem[\protect\citeauthoryear{MacKay}{MacKay}{2003}]{MacKay:2003}
MacKay D. J.~C.,  2003, Information Theory, Inference, and Learning Algorithms.
\url {http://www.inference.phy.cam.ac.uk/mackay/itila/}

\bibitem[\protect\citeauthoryear{{Poggio}, {Drimmel}, {Smart}, {Spagna}  \&
  {Lattanzi}}{{Poggio} et~al.}{2017}]{WarpGaiaDR1:2017}
{Poggio} E.,  {Drimmel} R.,  {Smart} R.~L.,  {Spagna} A.,   {Lattanzi} M.~G.,
  2017, \mn@doi [\aap] {10.1051/0004-6361/201629916}, \href
  {http://adsabs.harvard.edu/abs/2017A%26A...601A.115P} {601, A115}

\bibitem[\protect\citeauthoryear{{Reid} et~al.,}{{Reid}
  et~al.}{2014}]{Reid:2014}
{Reid} M.~J.,  et~al., 2014, \mn@doi [\apj] {10.1088/0004-637X/783/2/130},
  \href {http://adsabs.harvard.edu/abs/2014ApJ...783..130R} {783, 130}

\bibitem[\protect\citeauthoryear{{Reyl{\'e}}, {Marshall}, {Robin}  \&
  {Schultheis}}{{Reyl{\'e}} et~al.}{2009}]{Reyle:2009}
{Reyl{\'e}} C.,  {Marshall} D.~J.,  {Robin} A.~C.,   {Schultheis} M.,  2009,
  \mn@doi [\aap] {10.1051/0004-6361/200811341}, \href
  {http://adsabs.harvard.edu/abs/2009A%26A...495..819R} {495, 819}

\bibitem[\protect\citeauthoryear{{Robin}, {Reyl{\'e}}  \& {Marshall}}{{Robin}
  et~al.}{2008}]{Robin:2008}
{Robin} A.~C.,  {Reyl{\'e}} C.,   {Marshall} D.~J.,  2008, \mn@doi
  [Astronomische Nachrichten] {10.1002/asna.200811084}, \href
  {http://adsabs.harvard.edu/abs/2008AN....329.1012R} {329, 1012}

\bibitem[\protect\citeauthoryear{{Sch{\"o}nrich} \& {Aumer}}{{Sch{\"o}nrich} \&
  {Aumer}}{2017}]{Schoenrich:2017}
{Sch{\"o}nrich} R.,  {Aumer} M.,  2017, \mn@doi [\mnras]
  {10.1093/mnras/stx2189}, \href
  {http://adsabs.harvard.edu/abs/2017MNRAS.472.3979S} {472, 3979}

\bibitem[\protect\citeauthoryear{{Sch{\"o}nrich} \& {Dehnen}}{{Sch{\"o}nrich}
  \& {Dehnen}}{2018}]{Schoenrich:2018}
{Sch{\"o}nrich} R.,  {Dehnen} W.,  2018, \mn@doi [\mnras]
  {10.1093/mnras/sty1256}, \href
  {http://adsabs.harvard.edu/abs/2018MNRAS.478.3809S} {478, 3809}

\bibitem[\protect\citeauthoryear{{Sch{\"o}nrich}, {Binney}  \&
  {Dehnen}}{{Sch{\"o}nrich} et~al.}{2010}]{Schoenrich:2010}
{Sch{\"o}nrich} R.,  {Binney} J.,   {Dehnen} W.,  2010, \mn@doi [\mnras]
  {10.1111/j.1365-2966.2010.16253.x}, \href
  {http://adsabs.harvard.edu/abs/2010MNRAS.403.1829S} {403, 1829}

\bibitem[\protect\citeauthoryear{{Seabroke} \& {Gilmore}}{{Seabroke} \&
  {Gilmore}}{2007}]{Seabroke:2007}
{Seabroke} G.~M.,  {Gilmore} G.,  2007, \mn@doi [\mnras]
  {10.1111/j.1365-2966.2007.12210.x}, \href
  {http://adsabs.harvard.edu/abs/2007MNRAS.380.1348S} {380, 1348}

\bibitem[\protect\citeauthoryear{{Sellwood}}{{Sellwood}}{2013}]{Sellwood:2013}
{Sellwood} J.~A.,  2013, {Dynamics of Disks and Warps}.
p.~923, \mn@doi{10.1007/978-94-007-5612-0_18}

\bibitem[\protect\citeauthoryear{{Skrutskie} et~al.,}{{Skrutskie}
  et~al.}{2006}]{2MASS:2006}
{Skrutskie} M.~F.,  et~al., 2006, \mn@doi [\aj] {10.1086/498708}, \href
  {http://adsabs.harvard.edu/abs/2006AJ....131.1163S} {131, 1163}

\bibitem[\protect\citeauthoryear{{Smart}, {Drimmel}, {Lattanzi}  \&
  {Binney}}{{Smart} et~al.}{1998}]{Smart:1998}
{Smart} R.~L.,  {Drimmel} R.,  {Lattanzi} M.~G.,   {Binney} J.~J.,  1998,
  \mn@doi [\nat] {10.1038/33096}, \href
  {http://adsabs.harvard.edu/abs/1998Natur.392..471S} {392, 471}

\bibitem[\protect\citeauthoryear{{Sparke} \& {Casertano}}{{Sparke} \&
  {Casertano}}{1988}]{Sparke:1988}
{Sparke} L.~S.,  {Casertano} S.,  1988, \mn@doi [\mnras]
  {10.1093/mnras/234.4.873}, \href
  {http://adsabs.harvard.edu/abs/1988MNRAS.234..873S} {234, 873}

\bibitem[\protect\citeauthoryear{{Tang}, {Bressan}, {Rosenfield}, {Slemer},
  {Marigo}, {Girardi}  \& {Bianchi}}{{Tang} et~al.}{2014}]{Tang:2014}
{Tang} J.,  {Bressan} A.,  {Rosenfield} P.,  {Slemer} A.,  {Marigo} P.,
  {Girardi} L.,   {Bianchi} L.,  2014, \mn@doi [\mnras]
  {10.1093/mnras/stu2029}, \href
  {http://adsabs.harvard.edu/abs/2014MNRAS.445.4287T} {445, 4287}

\bibitem[\protect\citeauthoryear{{Wright}, {Egan}, {Kraemer}  \&
  {Price}}{{Wright} et~al.}{2003}]{T2STC:2003}
{Wright} C.~O.,  {Egan} M.~P.,  {Kraemer} K.~E.,   {Price} S.~D.,  2003,
  \mn@doi [\aj] {10.1086/345511}, \href
  {http://adsabs.harvard.edu/abs/2003AJ....125..359W} {125, 359}

\makeatother
\end{thebibliography}








\bsp	
\label{lastpage}
\end{document}